\documentclass{ws-ijmpe}
\newcommand{\am }{{\footnotesize AMPT }}
\newcommand{\amp}{{\footnotesize AMPT}}

\begin{document}
\markboth{Shakeel Ahmad et al}{Event-by-event fluctuations$\ldots$  }

\catchline{}{}{}{}{}

\title{Multifractal Characteristics of Multiparticle Production in \\Heavy-Ion Collisions at SPS Energies}

\author{SHAISTA KHAN and SHAKEEL AHMAD}

\address{Department of Physics, Aligarh Muslim University\\
Aligarh 202002, India
\\
sa\_hep@yahoo.com}



\maketitle

\begin{history}
\received{(received date)}
\revised{(revised date)}
\end{history}

\begin{abstract}
\noindent Entropy, dimensions and other multifractal characteristics of multiplicity distributions of relativistic charged hadrons produced in ion-ion collisions at SPS energies are investigated. The analysis of the experimental data is carried out in terms of phase space bin-size dependence of multiplicity distributions following the Takagi's approach. Yet another method is also followed to study the multifractality which, is not related to the bin-width and (or) the detector resolution, rather involves multiplicity distribution of charged particles in full phase space in terms of information entropy and its generalization, R\'{e}nyi's order-q information entropy. The findings reveal the presence of multifractal structure-- a remarkable property of the fluctuations. Nearly constant values of multifractal specific heat, 'c' estimated by the two different methods of analysis followed indicate that the parameter 'c' may be used as a universal characteristic of the particle production in high energy collisions. The results obtained from the analysis of the experimental data agree well with the predictions of Monte Carlo model \amp.
\end{abstract}

\section{Introduction}
\noindent Large particle density fluctuations in narrow phase space bins observed in cosmic ray (JACEE events)\cite{1} and accelerator experiments\cite{2} have generated considerable interest in the studies involving non-statistical fluctuations and the role of possible multifractal structure behind them\cite{3}. The method of scaled factorial moments (SFMs) proposed by Bialas and Peschanski\cite{4} has been extensively used to search for such phenomena in hadronic and heavy-ion collisions at widely different energies\cite{5,6,7,8,9,10,11,12,13,14,15,16,17}. The findings indicate that the presence of large density fluctuations in narrow pseudorapidity bins might be rare but not impossible. It has been reported\cite{4,5} that the observed large density fluctuations exhibit self similar properties as the bin size decreases up to the experimental/statistical limit. The fluctuation, if not appearing due to statistical reasons are envisaged to be the result of dynamical correlations amongst the produced particles that might arise due to the phase transitions from QGP to normal hadronic matter\cite{12}. After the observation of intermittency in e$^{+}$e$^{-}$, hadron-hadron (hh), hadron-nucleus (hA) and nucleus-nucleus (AA) collisions\cite{18,19,20}, numerous models were proposed to account for the intermittency or the power law behaviours of the SFMs\cite{4,5,21,22}, i.e.,\\
\begin{equation}
              F_{q} \propto (\delta y)^{-\alpha_{q}}   \hspace{2ex} \delta y \rightarrow 0 
\end{equation}

\noindent where exponent $\alpha_{q}$ increases with order q of the moment. \\

\noindent To account for the mathematical limit given by Eq.1, Takagi\cite{23} proposed a new simple method to examine the multifractal structure of multiplicity distributions (MDs).\\
 
\noindent It has been reported that constant specific heat (C.S.H) approximation\cite{24,25,26} widely used in standard thermodynamics, is applicable to the multifractal data too. Nearly constant values of multifractal specific heat have been observed from the analysis of the experimental data on hh and AA collisions\cite{12,27,28,29,30,31}; the analysis has been carried out following the method proposed by Takagi\cite{23}.\\

\noindent Fluctuations in physical observables in AA collisions are believed to be one of the important signals for the QGP formation because of the fact that in many body systems, phase transition would cause significant changes in quantum fluctuations of an observable from its average behaviour\cite{32,33}. When a system undergoes a phase transition, energy density remains the smooth function of temperature while the heat capacity changes abruptly\cite{32,34,35,36,37}. Entropy is taken yet another important characteristic of the system with many degrees of freedom\cite{32,38,39,40}. It has been pointed out\cite{32,34,41} that the systematic measurement of local entropy produced in AA collisions may provide direct information about the internal degrees of freedom of the QGP medium and its evolution. Bialas and Czyz\cite{38} proposed that R\'{e}nyi entropies may be used as a tool for studying the dynamical systems and are closely related to the thermodynamic entropy of the system -- the Shannon entropy. Furthermore, the generalisation of R\'{e}nyi's order-q information entropy contains information on the multiplicity moments and would serve as a useful tool for examining the multifractal characteristics of particle production\cite{26,42,43}. The advantage of this method of multifractal studies is that it is not related to the phase space bin widths and (or) the detector resolution but to the collision energy\cite{42}. An attempt is, therefore, made to carry out a well focussed study of the multifractal characteristics of multiparticle production, e.g. dimensions, entropy, etc., by analysing the experimental data on AA collisions at SPS energies following the Takagi's approach as well as by studying MDs and entropy.\\


\section{Details of the data}
\noindent Three samples of events produced in $^{16}$O-AgBr, $^{32}$S-AgBr and $^{32}$S-Gold interactions in emulsion at 200A GeV/c have been used. The number of events in these samples are 223, 452 and 542 respectively. These events are taken from the series of experiments carried out by EMU01 collaboration\cite{6,44,45,46}. All the relevant details of the data, like, selection criteria of events, track classification, extraction of AgBr group of events, method of measurements of emission angle, $\theta$ of relativistic charged particles, etc., may be found elsewhere\cite{12,32,34,44,47,48,49}. It may be mentioned that conventional emulsion techniques have two main advantages over the other detectors: (i) its 4$\pi$ solid angle coverage and (ii) data are free from biases due to full phase space acceptance. In the case of other detectors, only a fraction of charged particles are recorded due to the limited acceptance cone. This not only reduces the multiplicity but would also cause distortion of some of the event characteristics like particle density fluctuations\cite{34,49,50}. In order to compare the findings of the present work with the predictions of Monte Carlo model, \amp\cite{51}, event samples matching the experimental data are simulated using the code \amp-v-1.21-2.21. The events are simulated by taking into account the percentage of interactions which occurs in the collision of projectile with various target nuclei in emulsion  constituting the AgBr group of nuclei\cite{44,46,48}. The values of impact parameters, while generating events corresponding to each real data sets are so set that the mean multiplicity of the relativistic charged particles matches with those obtained for the experimental data sample.


\section{Formalism}
As mentioned earlier, the present method is carried out by following the Takaji's\cite{23} approach. a detailed description of bout the method of analysis has already been presented in refs.12 \& 23. Adopting this approach, the multifractal nature of multiparticle production may be examined by studying the variations of \(\langle n^{q}\rangle\) on \(\langle n\rangle\) on log-log plots and comparing with the following functional form:
\begin{eqnarray}
          ln\langle n^{q}\rangle  = A_{q} + k_{q}ln\langle n\rangle           
\end{eqnarray}
where
\begin{eqnarray}
          k_{q} = (q-1)D_{q} + 1          
\end{eqnarray}
gives the slope. The parameter \(A_q\) is a constant independent of bin-width.  \(\langle n\rangle\) denotes the mean multiplicity of charged particles in a particular bin, while \(D_q\) represents the generalized dimensions of order \(q\).\\

\noindent If a linear variation of $ln\langle n^{q}\rangle$ with $ln\langle n\rangle$ is observed, it would be an indication for the presence of dynamical fluctuations and a self similar process for multiparticle production. The slopes of the plots $ln\langle n^{q}\rangle$ vs $ln\langle n\rangle$ would give generalized dimensions D$_{q}$ for q $\ge$ 2 whereas the information dimension, D$_{1}$ may be estimated using the relation.\\
\begin{eqnarray}
          \frac{\langle nlnn\rangle}{\langle n\rangle } = C_{1} + D_{1}ln\langle n\rangle           
\end{eqnarray}

\noindent A decreasing trend in the values of D$_{q}$ with increasing q would indicate the presence of multifractality in the data.


\section{Results and Discussion}
\noindent The single particle distribution in pseudorapidity ($\eta$) distribution is non-flat. In order to reduce the effect of non-flatness, $\eta$ values of each particle are transformed into the cumulative variable X($\eta$) by using the following relation proposed by Bialas and Gazdzicki\cite{52}:
 
\begin{eqnarray}
           X(\eta) =  \frac{\int\limits_{\eta_{min}}^{\eta} \rho(\eta) d\eta}{\int\limits_{\eta_{min}}^{\eta_{max}} \rho(\eta) d\eta}   
\end{eqnarray}

\noindent where $\rho$($\eta$) denotes the single particle pseudorapidity density while $\eta_{min}$ and $\eta_{max}$ represent the minimum and maximum values of $\eta$-range considered. A relation similar to Eq.5 has been used to estimate X($\phi$) values too. \\

\noindent In order to have a flat single particle density distribution, $\eta$ and $\phi$ values of relativistic charged particles produced in each event are transformed into the cumulative variables X($\eta$) and X($\phi$) using Eq.5. Relativistic charged particles having their $\eta$ and $\phi$ values in the ranges, \(\eta_c-3.0 \leq \eta \leq \eta_c+3.0\)
 and \((0 \leq \phi \leq 2\pi)\) have been considered; $\eta_{c}$ represents the centre of $\eta$ distribution. Thus, in X-space values of variables $\eta$ and $\phi$ would lie in the range 0-1 and have a flat distribution.\\

\noindent Values of $\langle n\rangle $, $\langle n^{q}\rangle $ and $\langle nlnn\rangle $ are calculated by dividing X($\eta$) and X($\phi$) space into M bins of equal width and by varying M from 2 to 30 in steps of 2. Variations of $ln\langle n^{q}\rangle$ and $\langle nlnn\rangle $ with $ln\langle n\rangle$ in $\eta$ and $\phi$  spaces for the three data sets are presented in Figs.1-2. Results from the analysis of \am samples are also shown in the same figures. It is evident from the figures that values of $ln\langle n^{q}\rangle$ and $\langle nlnn\rangle $ increase with $ln\langle n\rangle$. It may, further, be noted that experimental data are nicely reproduced by the \am model. The values of slope k$_{q}$ and the generalized dimensions D$_{q}$ for q = 2-6 are estimated using Eqs.2 and 3. The values of information dimension D$_{1}$ are calculated using Eq.4. These values for $\eta$ and $\phi$ spaces are presented in Table~1. Since the Takagi's multifractal moments are observed to show a power law behaviour, it may be remarked that the data exhibits the self-similar property of multiparticle production process.\\

\noindent It may be noted from Table~1 that values of D$_{q}$ decreases with increasing order q. Such a decreasing trend of D$_{q}$ with q can be related to the scaling behaviour of q point correlation integrals\cite{26}. Thus, the observed behaviour of D$_{q}$ against q also indicates the multifractal nature of MDs in full phase space in ion-ion collisions at SPS energies. The presence of multifractality, although predicts D$_{q}$ to decrease with q, yet no further useful information about the q-dependence of D$_{q}$ spectrum can be extracted from which conclusions about the scaling properties of q-correlation integrals\cite{5} may be drawn. It has been suggested \cite{26,53} that in constant specific heat approximation, D$_{q}$ dependence on q would acquire the following simple form:

\begin{eqnarray}
D_{q} \simeq (a-c)+c\frac{ln(q)}{(q-1)}
\end{eqnarray} 

\noindent where 'a' is the information dimension, $D_{1}$, while 'c' denotes the multifractal specific heat. The linear trend of variations $D_{q}$ with $ln$(q)/(q-1), given by Eq.6, is expected to be observed for multifractals. On the basis of classical analogy with specific heat of gases and solids the value of 'c' is predicted\cite{54} to be independent of temperature in a wide range of q. In order to test the validity of Eq.6, values of $D_{q}$ against $ln$(q)/(q-1) for the three data sets (real and \amp) are plotted in Figs.3 and 4. A linear rise of $D_{q}$ with $ln$(q)/(q-1) is observed in the figures for all the data sets considered. The lines in the figures represent the best fits to the data obtained using Eq.6. The values of the intercepts, (a - c) and slopes, 'c', are listed in Table~2. It may be noted from the table that the values of multifractal specific heat, 'c' for all the data sets are nearly the same $\sim$ 0.2 indicating its independence on the beam/target mass.  It is also reflected from the table that the experimental values of 'c' are close to corresponding \am predicted values. The values of the intercept are, however, noted to be somewhat different for different data sets. \\

\noindent As mentioned earlier entropy is regarded yet another important characteristic of the system with many degrees of freedom\cite{34,38,39}. Processes in which particles are produced may be considered as the one so called dynamical systems\cite{32,38,39,40} in which entropy is generally produced. Systematic measurements of local entropy produced in relativistic AA collisions may provide direct information about the internal degrees of freedom of QGP medium and its evolution\cite{32,34,41}. It has been pointed out\cite{55} that in high energy collisions particle production occurs on the maximum stochasticity, i.e., they should follow the maximum entropy principle\cite{56}.\\

\noindent This type of stochasticity may also be quantified in terms of information entropy which is more natural and general parameter to measure the chaoticity in branching processes\cite{56}. Simak et al\cite{57}, by using the information entropy showed that MDs of charged particles produced in limited and full phase space in hadron-hadron (hh) collisions exhibit a new type of scaling law in the energy range $\sqrt{s}$ $\sim$ 19 to 900 GeV. Analyses of experimental data on hh\cite{41,57,58,59} and AA\cite{32,34,60,61} collisions carried out by several workers reveal that entropy per unit rapidity appears to be an energy independent quantity, suggesting the presence of entropy scaling. Bialas and Czyz\cite{38,62,63,64} proposed that R\'{e}nyi entropies may also be used as a tool for studying the dynamical systems and are closely related to the thermodynamic entropy of the system -the Shannon entropy. It has also been pointed out\cite{65} that R\'{e}nyi entropies are worthy of investigation as they represent various interesting properties and may serve as useful tool for examining the correlations amongst the particles produced in high energy collisions\cite{64}.\\

\noindent The entropy of the charged particle MDs, Shannon's information entropy is defined as\cite{32,57};
\begin{eqnarray}
              S=-\sum P_{n}lnP_{n},  \hspace{2ex} \sum P_{n} = 1
\end{eqnarray}

\noindent where n is the charged particle multiplicity of an event while P$_{n}$ denotes the probability of production of n charged particles. R\'{e}nyi's order q information entropy, which is generalization of Shannon's entropy is given by\cite{26,43,66} the relation 

\begin{eqnarray}
          I_{q}=\frac{1}{q-1}ln\sum P_{n}^{q} \hspace{2ex} for \hspace{2ex} q \neq 1          
\end{eqnarray}

\noindent where, for q = 1  $\displaystyle{\lim_{q \to 1} I_{q} = I_{1} = S}$ 

\noindent The generalized dimensions of order q may be evaluated as 
\begin{eqnarray}
D_{q} = I_{q}/Y_{m}
\end{eqnarray}
 
\begin{eqnarray}
where \hspace{2ex}  Y_{m} = ln[(\sqrt{s}-2m_{n}\langle n_{p}\rangle)/m_{\pi}] = ln n_{max} 
\end{eqnarray}

\noindent The quantity Y$_{m}$ in Eq.10 denotes the maximum rapidity, $\sqrt{s}$ is the centre-of-mass energy, m$_{\pi}$ is the pion rest mass,  n$_{max}$ represents the maximum multiplicity of relativistic charged particles produced in a collision and $\langle n_{p}\rangle$ denotes the mean number of participating nucleons. \\

\noindent It is evident from Eqs.7-9 that for a given q, (I$_{q}$)$_{max}$ = $ln$n$_{max}$. The maximum entropy is achieved for the greatest "chaos" of a uniformly distributed probability function P$_{n}$ = 1/n$_{max}$\cite{43}. Thus, from Eq.8 one gets D$_{q}$ = I$_{q}$/(I$_{q}$)$_{max}$.\\

\noindent Values of generalized dimensions D$_{q}$ for q = 2-6 are calculated for the real and simulated data sets using Eqs.7-9 and are presented in Table~3. It may be noted from the table that values of D$_{q}$ for various data sets are somewhat higher as compared to those obtained using Takagi's approach (Table~1). In order to estimate the values of multifractal specific heat, variations of D$_{q}$ with $ln$(q)/(q-1) are plotted in Fig.5. The lines in the figure represent the best fits to the data obtained using Eq.6. The values of slopes(the multifractal specific heat) and the intercepts(a - c) are listed in Table~4. It may be noticed from the table that the values of 'c' obtained following the Takagi's method match with those estimated from entropy studies. The values of D$_{1}$ estimated from the two approaches are also close to each other.\\

\noindent It may be remarked here that the values of multifractal specific heat estimated in the present study are close to those obtained by us\cite{12} by analysing the data on 14.5A GeV/c Si-AgBr and 10.6A GeV/c Au-AgBr collisions following the Takagi's approach\cite{23}. Incidentally similar values of 'c' have been reported by Bershadskii\cite{24} for 10.6A GeV/c Au-nucleus collisions. In proton-proton interactions too, the value of multifractal specific heat has been reported\cite{24,43,53} to be $\sim 0.25$ in the energy range $\sim$ 200-800 GeV. These findings, therefore, tend to suggest that the constant specific heat approximation is applicable to the multiparticle production in relativistic hh, hA and AA collisions. Furthermore, nearly the same values of multifractal specific heat observed in the present study as well as the ones reported by several workers earlier does suggest that the parameter 'c' may be taken as a universal characteristics of relativistic hadronic and ion-ion collisions.\\

\section{Conclusions}
\noindent Based on the findings of the present work the following conclusions may be arrived at:
\begin{enumerate}
\item There are reasonably good indications of the presence of multifractal structure in the MDs of relativistic charged particles in ion-ion collisions at SPS energies.
\item Nearly the same values of multifractal specific heat are found by estimating the values of generalized dimensions, D$_{q}$ and examining the dependence of D$_{q}$ on ln(q)/(q-1) in both $\eta$ and $\phi$ spaces.
\item Values of D$_{q}$ and multifractal specific heat, 'c', evaluated from R\'{e}nyi's information entropy are found to be in good accord with those obtained by following the Takagi's approach.
\item Results from the analysis of \am data are observed to match with those obtained from the experimental data. 
\end{enumerate}

\newpage
\begin{figure}[th]
\centerline{\psfig{file=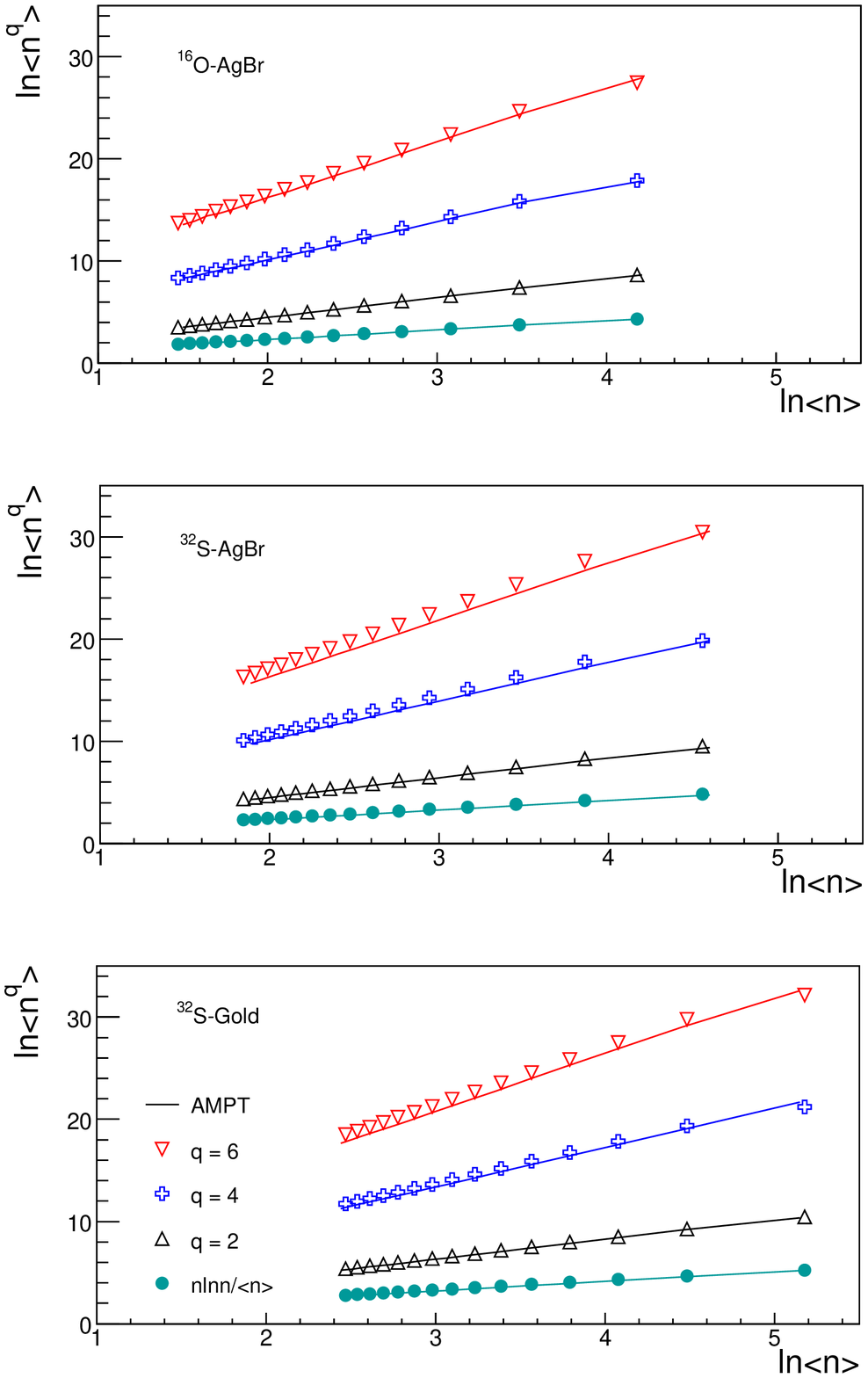, width=10cm}}
\vspace*{8pt}
\caption{Variations of $ln\langle n^{q}\rangle$ with $\langle nlnn\rangle $ in $\eta$-space for the relativistic charged  particles produced in $^{16}$O-AgBr, $^{32}$S-AgBr and $^{32}$S-Gold interactions at 200 A GeV/c.}
\end{figure}

\newpage
\begin{figure}[th]
\centerline{\psfig{file=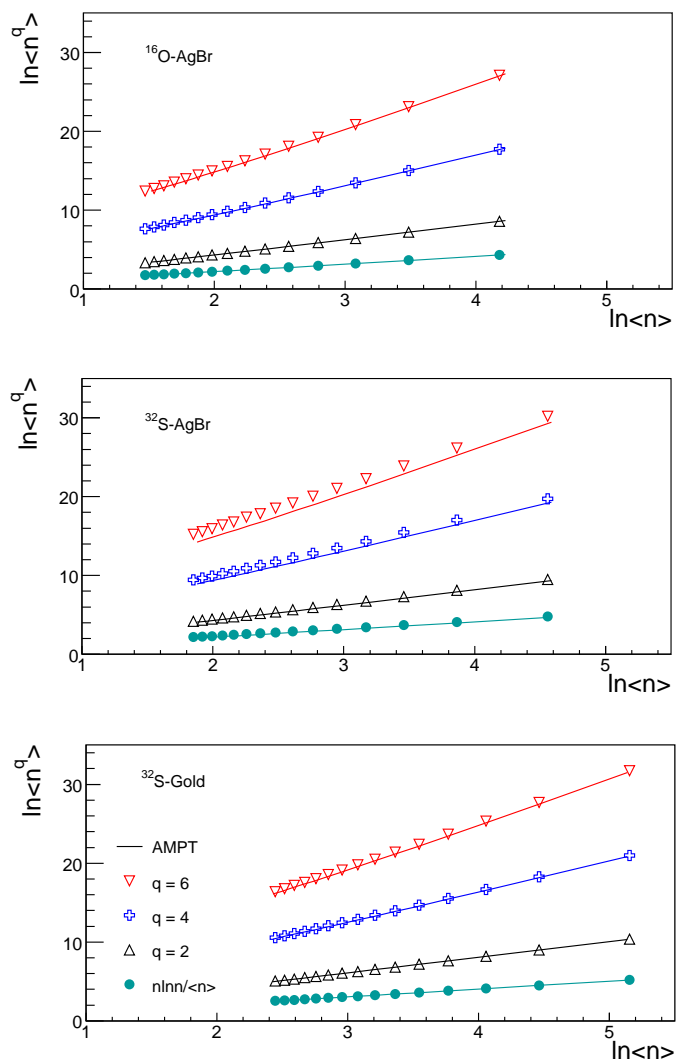,width=10cm}}
\vspace*{8pt}
\caption{The same plot as in Fig.1 but in $\phi$-space.}
\end{figure}
\newpage
\begin{figure}[th]
\centerline{\psfig{file=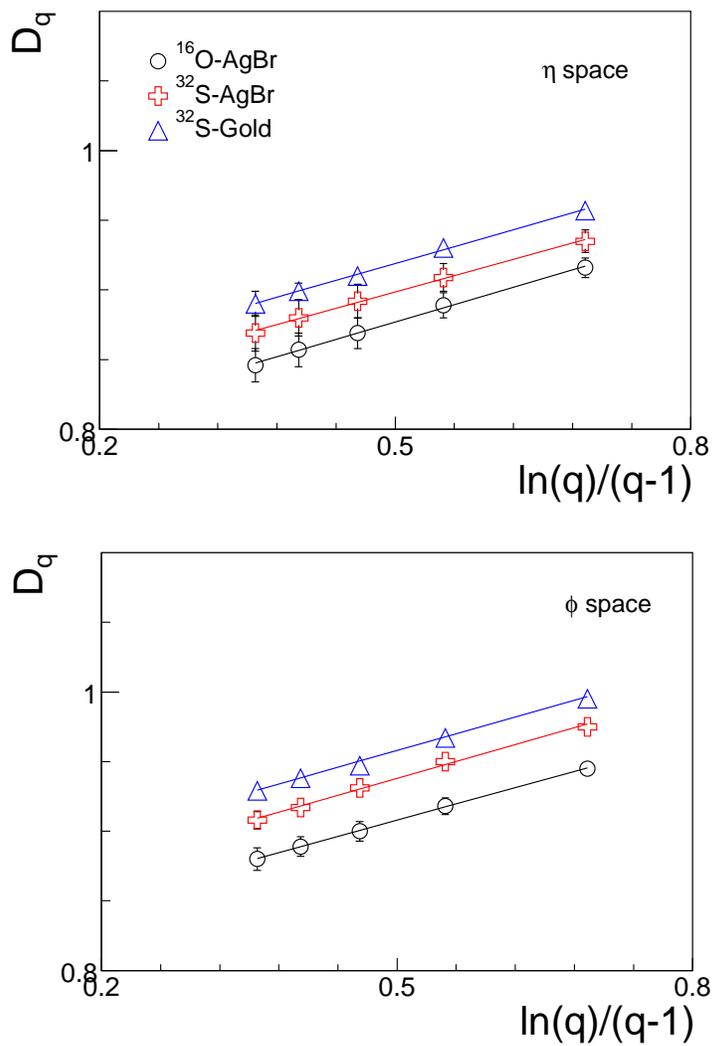,width=10cm}}
\vspace*{8pt}
\caption{Dependence of D$_{q}$ on ln(q)/(q-1) for the experimental events in $\eta$ and $\phi$ space.}
\end{figure}
\newpage
\begin{figure}[th]
\centerline{\psfig{file=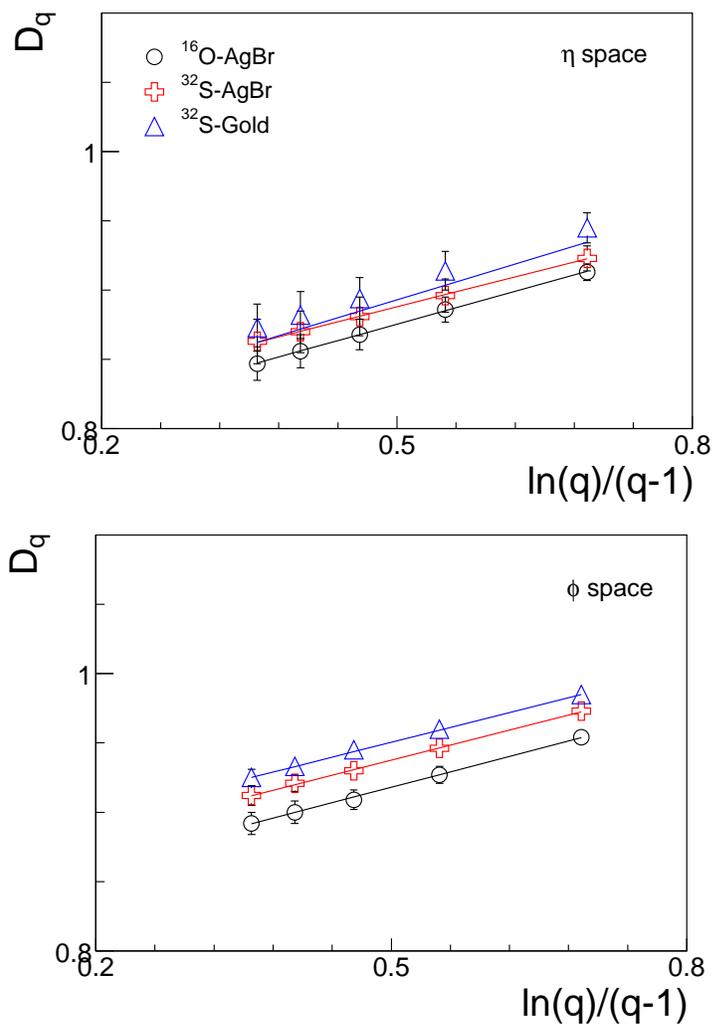,width=10cm}}
\vspace*{8pt}
\caption{The same plot as in Fig.3 but for the \am events.}
\end{figure}
\newpage
\begin{figure}[th]
\centerline{\psfig{file=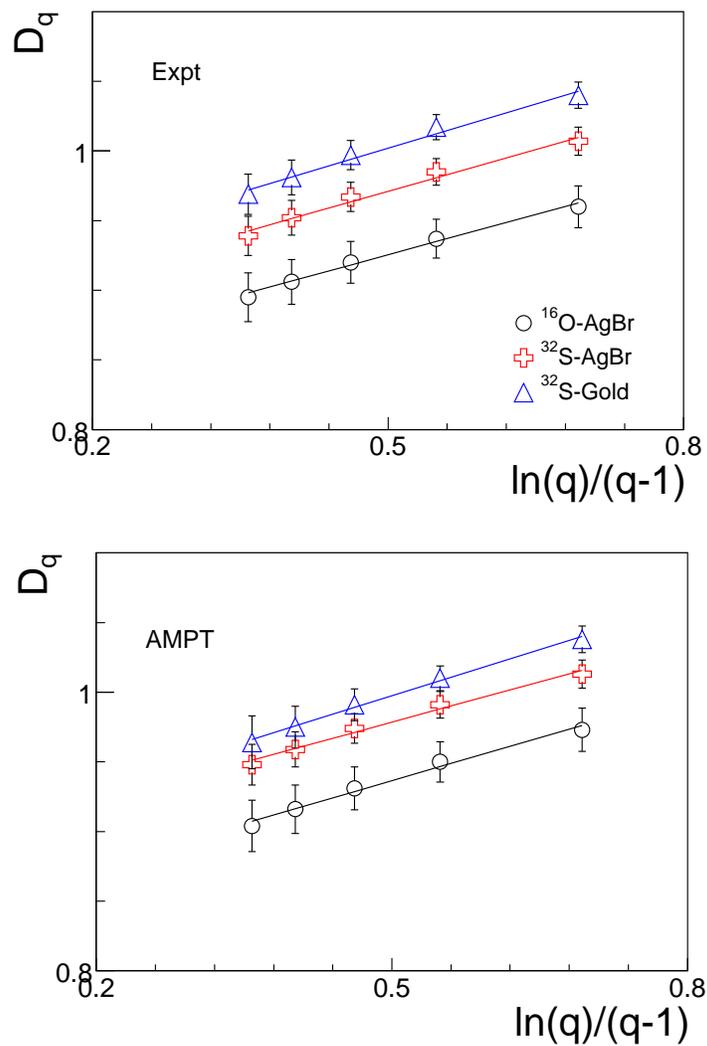,width=10cm}}
\vspace*{8pt}
\caption{D$_{q}$ vs \(ln\)(q)/(q-1) plots for the experimental and simulated data at the three energies; Dq values shown are obtained using Eqs.7-9.}
\end{figure}

\newpage
\begin{footnotesize}
\begin{table}[htbp]
 \noindent Table~1: Values of D$_{q}$ for $\eta$ and $\phi$ spaces for the real and simulated data sets: Dq values for q = 2-6 are calculated using Eqs.2 and 3, while that for q = 1 are obtained using Eq.4.
\begin{center}
\resizebox{11.8cm}{3.6cm}{
\begin{tabular}{|c|c|c|c|c|c|c|} \hline
q &  1 & 2 & 3 & 4 & 5 & 6 \\ \hline
&           \multicolumn{6}{|c|}{$\eta$-space} \\ [2mm] \hline
$^{16}$O-AgBr   &  0.938 $\pm$ 0.005  &  0.928 $\pm$ 0.005  &  0.900 $\pm$ 0.006  & 0.880 $\pm$ 0.007  &  0.866 $\pm$ 0.008  &  0.854 $\pm$ 0.009   \\[2mm] 
AMPT            &  0.942 $\pm$ 0.005  &  0.929 $\pm$ 0.006  &  0.898 $\pm$ 0.008  & 0.879 $\pm$ 0.009  &  0.865 $\pm$ 0.010  &  0.855 $\pm$ 0.011   \\[2mm] \hline
$^{32}$S-AgBr   &  0.978 $\pm$ 0.003  &  0.965 $\pm$ 0.003  &  0.942 $\pm$ 0.004  & 0.926 $\pm$ 0.005  &  0.913 $\pm$ 0.005  &  0.904 $\pm$ 0.006 \\[2mm] 
AMPT            &  0.969 $\pm$ 0.003  &  0.955 $\pm$ 0.005  &  0.924 $\pm$ 0.006  & 0.909 $\pm$ 0.007  &  0.897 $\pm$ 0.008  &  0.888 $\pm$ 0.009   \\[2mm] \hline
$^{32}$S-Gold   &  0.990 $\pm$ 0.001  &  0.970 $\pm$ 0.002  &  0.947 $\pm$ 0.008  & 0.930 $\pm$ 0.005  &  0.918 $\pm$ 0.007  &  0.906 $\pm$ 0.010   \\[2mm] 
AMPT            &  0.975 $\pm$ 0.002  &  0.972 $\pm$ 0.004  &  0.945 $\pm$ 0.006  & 0.926 $\pm$ 0.007  &  0.914 $\pm$ 0.008  &  0.903 $\pm$ 0.009   \\ \hline 
                &        \multicolumn{6}{|c|}{$\phi$-space} \\ [2mm] \hline 
$^{16}$O-AgBr   &  0.938 $\pm$ 0.004  &  0.923 $\pm$ 0.005  &  0.898 $\pm$ 0.006  & 0.880 $\pm$ 0.006  &  0.867 $\pm$ 0.007  &  0.855 $\pm$ 0.008   \\[2mm] 
AMPT            &  0.960 $\pm$ 0.004  &  0.959 $\pm$ 0.005  &  0.930 $\pm$ 0.006  & 0.914 $\pm$ 0.007  &  0.902 $\pm$ 0.007  &  0.895 $\pm$ 0.008   \\[2mm] \hline
$^{32}$S-AgBr   &  0.954 $\pm$ 0.003  &  0.950 $\pm$ 0.003  &  0.924 $\pm$ 0.004  & 0.904 $\pm$ 0.004  &  0.890 $\pm$ 0.006  &  0.878 $\pm$ 0.010 \\[2mm] 
AMPT            &  0.972 $\pm$ 0.002  &  0.976 $\pm$ 0.004  &  0.948 $\pm$ 0.005  & 0.932 $\pm$ 0.005  &  0.920 $\pm$ 0.006  &  0.912 $\pm$ 0.006   \\[2mm] \hline
$^{32}$S-Gold   &  0.986 $\pm$ 0.001  &  0.979 $\pm$ 0.002  &  0.951 $\pm$ 0.003  & 0.936 $\pm$ 0.004  &  0.924 $\pm$ 0.004  &  0.916 $\pm$ 0.004   \\[2mm] 
AMPT            &  0.997 $\pm$ 0.002  &  0.994 $\pm$ 0.003  &  0.967 $\pm$ 0.004  & 0.949 $\pm$ 0.004  &  0.936 $\pm$ 0.005  &  0.928 $\pm$ 0.006   \\ \hline 
\end{tabular}
}
\end{center}
\end{table}
\end{footnotesize} 

\newpage
\begin{footnotesize}
\begin{table}[htbp]
	\noindent Table~2: Values of parameters (a-c) and 'c', appearing in Eq.6 for various event samples.
\begin{center}
 \begin{tabular}{|c|c|c|c|c|c|c|c|c|c|} \hline
Type of &  \multicolumn{2}{|c|}{Expt.} & \multicolumn{2}{|c|}{AMPT} \\ \cline{2-5}
interactions &    a-c &  c    &  a-c &  c \\[2mm] \hline
    &                          \multicolumn{4}{|c|}{$\eta$-space} \\ [2mm] \cline{1-5}
$^{16}$O-AgBr    &  0.777 $\pm$ 0.004  &  0.220 $\pm$ 0.007  &  0.776 $\pm$ 0.001  &  0.220 $\pm$ 0.002 \\[2mm] \hline
$^{32}$S-AgBr    &  0.840 $\pm$ 0.003  &  0.182 $\pm$ 0.007  &  0.817 $\pm$ 0.002  &  0.198 $\pm$ 0.004 \\[2mm] \hline
 $^{32}$S-Gold    &  0.841 $\pm$ 0.005  &  0.188 $\pm$ 0.009  &  0.831 $\pm$ 0.003  &  0.205 $\pm$ 0.005 \\[2mm] \hline
    &                          \multicolumn{4}{|c|}{$\phi$-space} \\ [2mm]  \cline{1-5}
 $^{16}$O-AgBr    &  0.786 $\pm$ 0.004  &  0.200 $\pm$ 0.009  &  0.825 $\pm$ 0.001  &  0.192 $\pm$ 0.003 \\[2mm] \hline
 $^{32}$S-AgBr    &  0.803 $\pm$ 0.005  &  0.214 $\pm$ 0.009  &  0.843 $\pm$ 0.000  &  0.191 $\pm$ 0.001 \\[2mm] \hline
 $^{32}$S-Gold    &  0.849 $\pm$ 0.001  &  0.187 $\pm$ 0.002  &  0.857 $\pm$ 0.001  &  0.199 $\pm$ 0.003 \\[2mm] \hline

\end{tabular}
\end{center}
\end{table}
\end{footnotesize}

\newpage
\begin{footnotesize}
\begin{table}[htbp]
 \noindent Table~3: Values of D$_{q}$ for q = 2-6 for the real and \am events at the three energies; D$_{q}$ values are estimated using Eqs.7-9.
\begin{center}
\resizebox{14.2cm}{2.2cm}{
\begin{tabular}{|c|c|c|c|c|c|} \hline
q  & 2 & 3 & 4 & 5 & 6 \\ \hline
$^{16}$O-AgBr  & 0.960 $\pm$ 0.030 & 0.937 $\pm$ 0.028 & 0.920 $\pm$ 0.030 & 0.906 $\pm$ 0.032 & 0.895 $\pm$ 0.035 \\[2mm] 
       AMPT    & 0.973 $\pm$ 0.031 & 0.950 $\pm$ 0.029 & 0.931 $\pm$ 0.031 & 0.916 $\pm$ 0.035 & 0.904 $\pm$ 0.037 \\[2mm] \hline
$^{32}$SAgBr   & 1.007 $\pm$ 0.020 & 0.985 $\pm$ 0.019 & 0.967 $\pm$ 0.021 & 0.952 $\pm$ 0.025 & 0.939 $\pm$ 0.028 \\[2mm]
       AMPT    & 1.013 $\pm$ 0.020 & 0.991 $\pm$ 0.019 & 0.974 $\pm$ 0.021 & 0.959 $\pm$ 0.025 & 0.948 $\pm$ 0.029 \\[2mm] \hline
$^{32}$S-Gold  & 1.040 $\pm$ 0.019 & 1.017 $\pm$ 0.018 & 0.997 $\pm$ 0.021 & 0.981 $\pm$ 0.025 & 0.969 $\pm$ 0.029 \\[2mm]
       AMPT    & 1.038 $\pm$ 0.019 & 1.010 $\pm$ 0.018 & 0.991 $\pm$ 0.023 & 0.975 $\pm$ 0.030 & 0.964 $\pm$ 0.038 \\  \hline

\end{tabular}
}
\end{center}
\end{table}
\end{footnotesize}

\begin{footnotesize}
\begin{table}[htbp]
	\noindent Table~4: Values of intercepts (a-c) and slopes, 'c' for various data sets obtained (using Eq.6) from the best fits to the data plotted in Fig.5.
\begin{center}
 \begin{tabular}{|c|c|c|c|c|c|c|c|c|c|} \hline
Type of &  \multicolumn{2}{|c|}{Expt.} & \multicolumn{2}{|c|}{AMPT} \\ \cline{2-5}
interactions &    a-c &  c    &  a-c &  c \\[2mm] \hline
$^{16}$O-AgBr    &  0.829 $\pm$ 0.005  &  0.193 $\pm$ 0.011  &  0.834 $\pm$ 0.007  &  0.205 $\pm$ 0.013 \\[2mm] \hline
 $^{32}$S-AgBr    &  0.871 $\pm$ 0.008  &  0.201 $\pm$ 0.015  &  0.882 $\pm$ 0.006  &  0.193 $\pm$ 0.013 \\[2mm] \hline
  $^{32}$S-Gold    &  0.896 $\pm$ 0.008  &  0.212 $\pm$ 0.014  &  0.887 $\pm$ 0.004  &  0.221 $\pm$ 0.007 \\[2mm] \hline
 
\end{tabular}
\end{center}
\end{table}
\end{footnotesize} 
\end{document}